\documentclass[fleqn,usenatbib]{rasti}

\usepackage{newtxtext,newtxmath}

\usepackage[T1]{fontenc}
\DeclareRobustCommand{\VAN}[3]{#2}
\let\VANthebibliography\thebibliography
\def\thebibliography{\DeclareRobustCommand{\VAN}[3]{##3}\VANthebibliography}

\usepackage{graphicx}	
\usepackage{amsmath}	
\usepackage{subcaption}
\usepackage{placeins}
\usepackage{comment}

\newcommand{\vo}{\vec{o}\@ifnextchar{^}{\,}{}}

\title{Lord of the (sub-)Rings : Mapping the surface reflectance and spin-axis of Ajisai}

\author[R. J. S. Airey et al.]{
Robert J. S. Airey,$^{1,2}$\thanks{E-mail: Robert.Airey@warwick.ac.uk}
Paul Chote,$^{1,2}$
James A. Blake,$^{1,2}$
James McCormac,$^{1,2}$
Billy Shrive,$^{1,2}$
Don Pollacco,$^{1,2}$ \and
\space Benjamin F. Cooke$^{1,2}$
\\
$^{1}$Department of Physics, University of Warwick, Gibbet Hill Road, Coventry, CV4 7AL, UK\\
$^{2}$Centre for Space Domain Awareness, University of Warwick, Gibbet Hill Road, Coventry, CV4 7AL, UK\\
}

\date{Accepted 2025 November 20. Received 2025 October 16; in original form 2025 July 23}

\pubyear{\the\year{}}

\begin{document}
\label{firstpage}
\pagerange{\pageref{firstpage}--\pageref{lastpage}}
\maketitle

\begin{abstract}
Active debris removal techniques are posed to become an important tool in maintaining the safety of the near-Earth space environment. These techniques rely on a clear understanding of the rotational motion of the debris targets, which is challenging to constrain from unresolved imaging. The Ajisai satellite provides an ideal test case for developing and demonstrating these techniques due to its simple geometry and well constrained spin behaviour. We present four observations of the Ajisai satellite taken with SuperWASP in August of 2019, where high cadence photometry was extracted from streaked images as a part of a larger survey of Low Earth Orbit. We develop an MCMC-driven method to determine the spin-state of Ajisai by comparing the alignment between a map of modelled mirror positions and a novel derived map of surface reflectivity. We generally find good agreement within the expectation and uncertainties set by empirical models and our determined spin-state solutions align the surface reflectivity map and modelled mirror location well. Our results show that streak photometry can be used to recover the spin-axis and rotation period of fast-spinning objects such as Ajisai using modest ground-based instrumentation, making it readily scalable to a wider range of targets and observatories.
\end{abstract}




\section{Introduction}
The population of non-cooperative objects in the near-Earth environment has continued to grow in recent years \citep{blake2022looking,2024amos.conf..134S}, such that active debris removal (ADR) techniques to achieve a more sustainable space and reduce collision risk are being proactively proposed and discussed \citep[see e.g.][]{2022AdSpR..70.2976E,2024AcAau.225..676B,2024AcAau.225..891R,2024AcAau.220..108S}. In particular, most ADR techniques rely on a service vehicle to attach itself to the debris target; therefore, it is critical that the rotational state of the debris object be known.

Using photometric measurements with the aim of estimating the state (spin-axis and period) of a non-cooperative object has become an active area of research \citep[see e.g.][]{2018AdSpR..61..844S,blake2021debriswatch, soton457200,2024amos.conf..115M}. However, these techniques may rely on a BRDF (Bidirectional Reflectance Distribution Function) model \citep{2012tres.book.....H} being implemented to generate synthetic light curves which can be compared to truth observed light curves. Although, more recent methods have been developed which can infer spin properties directly from temporal frequency changes without assuming a particular BRDF \citep[see e.g.][]{2020AdSpR..65.1518Z, 2024JAnSc..71...41B}.

This study aims to determine the attitude state of the Ajisai satellite purely from TLE-derived vectors and high-cadence optical photometry, expanding on the previous work by \citet{2006amos.confE..76H} (Epoch-method) and, more recently, by \citet{2024AdSpR..74.5725K} (photometric-patterns method). It is important to note that Ajisai’s spin state has been previously investigated using different approaches. Notably, \cite{CALATRONI2025} employed high-frequency optical photometry (5–10 kHz) to determine the full rotation state—including the spin axis, period, and initial rotation phase—with high single-pass accuracy. \cite{2016AdSpR..57..983K} achieved long-term monitoring of Ajisai’s attitude evolution using 2 kHz Satellite Laser Ranging (SLR) data which benefits from much broader temporal coverage at a high precision. Earlier optical studies such as \cite{2017AdSpR..60.1389K}, demonstrated that attitude determination can also be performed with modest ground-based instrumentation, and therefore this work is particularly relevant as a precursor to the present study. The reflectivity of the Ajisai mirrors have also been studied in the works of \cite{2019AdSpR..64..957K} and \cite{2020AcAau.174...24K} in which the focus is purely on studying the mirror reflectivity mapping as opposed to attitude determination.

Observational cadences on the order of milliseconds are needed to ascertain the spin-states of the fastest spinning objects \citep{2019LPICo2109.6075K}, which places strict limits on observing cadence with traditional CCD observations due to their relatively slow readout times and inefficiency at capturing rapid exposures without introducing significant dead time. Here we present a technique that doesn't rely on specialized kHz-level instrumentation, making it more readily scalable to a wider range of targets and observatories.

\subsection{Ajisai Background}
\begin{figure}
    \centering
    \includegraphics[height = 0.4\textheight]{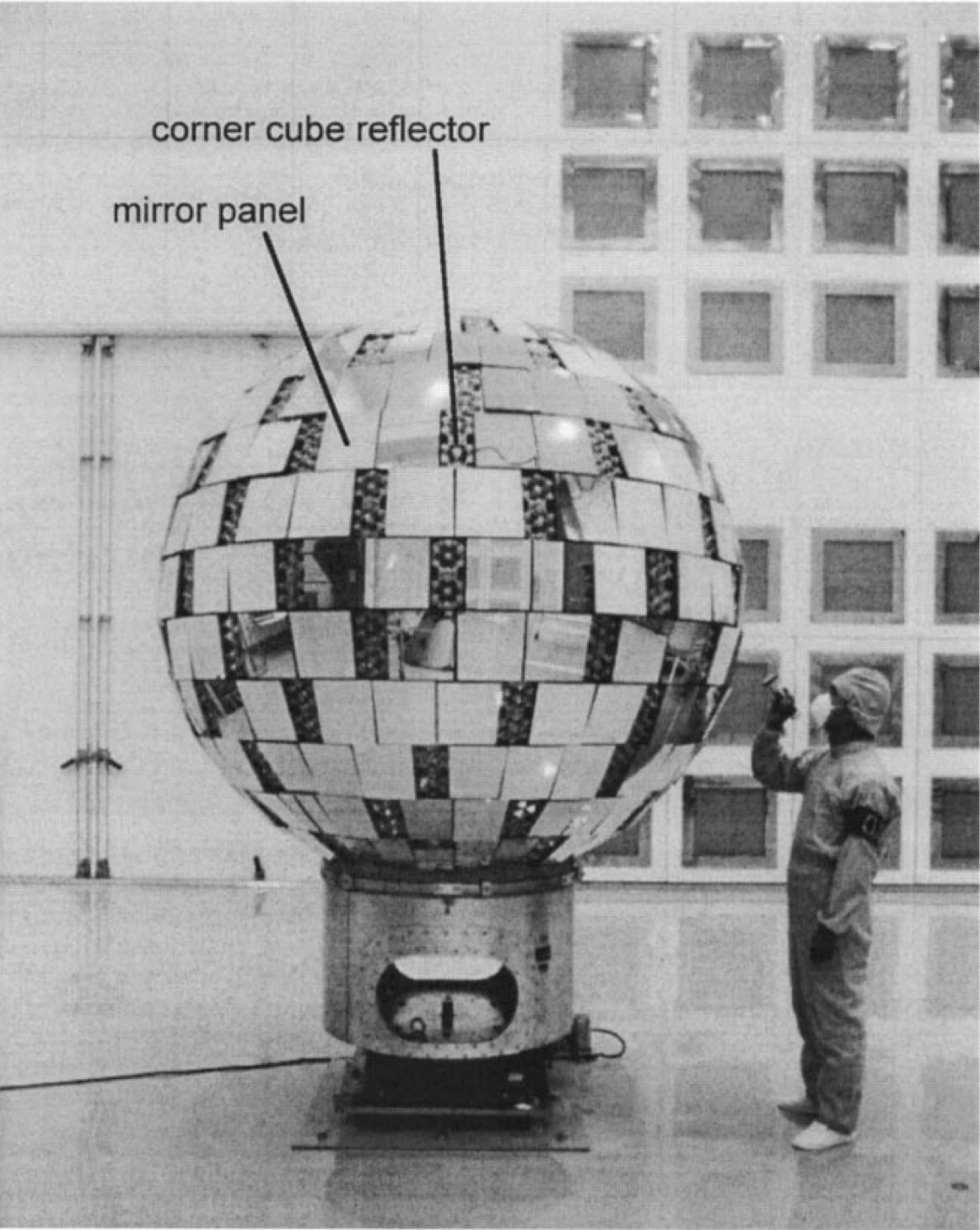}
    \caption{The Ajisai satellite with mirror panels and CCRs (corner cube reflectors) covering its spherical surface labelled on the figure. This figure is taken from \protect\cite{2000ITGRS..38.1417O}}
    \label{fig:ajisai_scale}
\end{figure}
The Ajisai satellite (NORAD: 16908, COSPAR: 1986-061A) was launched in 1986 as part of a geodetic mission sponsored by the NASDA (National Space Development Agency of Japan), with the goal of improving the accuracy of domestic geodetic triangulation with respect to other networks and accurately measuring the position of the remote Japanese islands \citep{UNnotes}. The satellite is in Low Earth Orbit (LEO) approximately 1500 km above the Earth's surface with an orbit inclined by approximately 50$^\circ$. Ajisai is a 685 kg hollow sphere with a diameter of 2.15 meters, and the surface is covered with 318 mirrors and 1436 corner-cube retroreflectors (CCRs)\citep{1987ITGRS..25..526S} (see Figure \ref{fig:ajisai_scale}). Work done in \cite{Ajisai_spin_param} determined that Ajisai is spinning in the clockwise direction. The mirrors of Ajisai are curved at a radius of 8.55 m ($\pm$ 0.15) and have a surface area of up to 393 cm$^2$ each \citep{2019AdSpR..64..957K}. There are two modes of operation that enable precise geodetic measurements: one involves sending a ground-based laser beam to the satellite’s CCRs and measuring the round-trip travel time (laser ranging); the other involves imaging the illuminated satellite from ground stations against background stars to determine its position \citep{eoportal_egs}.

The mirrors covering the surface are organized into 12 primary latitudinal rings, along with two polar caps. Each primary ring is composed of several sub-rings, with each sub-ring containing three mirrors positioned at the same latitude and inclined in the same latitudinal direction. This means that within a given revolution of the object around its spin-axis, a sub-ring of mirrors should always be able to reflect light towards an observer.
\begin{figure*}
    \centering\includegraphics{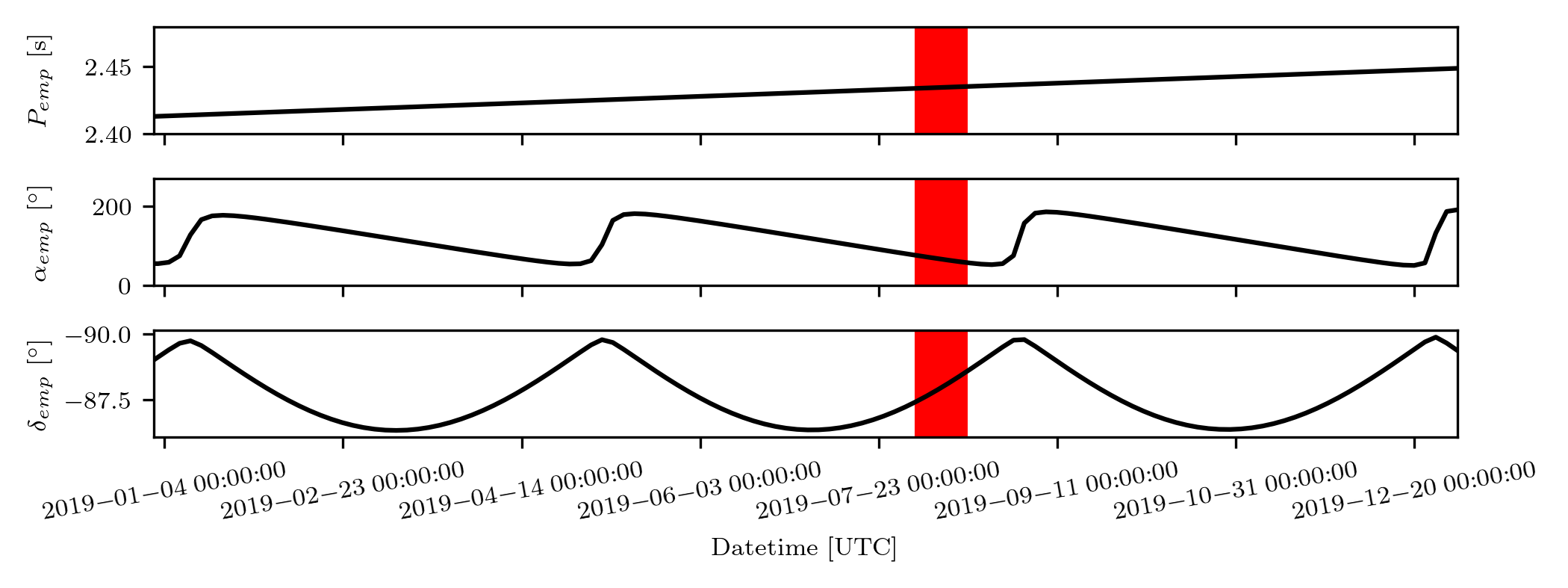}
    \caption{Propagation of the empirical model \protect\citep{2016AdSpR..57..983K} for the orientation and period of Ajisai's spin axis. The red segment corresponds to the extent of the dates in which Ajisai was observed with SuperWASP. The empirical model defines $\delta_{emp}$ as being measured relative to the south celestial pole, with a clockwise spin relative to this axis.}
    \label{fig:ajisai_empirical}
\end{figure*}
Ajisai was spin-stabilised, with an initial spin vector aligned with the Earth's rotation and a period of 1.5 s \citep{1987ITGRS..25..526S}. The spin-axis in the International Celestial Reference Frame (ICRF) and period of Ajisai at any time of observation can be approximated using the empirical models from \cite{2016AdSpR..57..983K}. These are a set of time-dependent equations which describe the expected evolution of the precession and nutation cones of the spin-axis, and the period as an exponential trend function. The expected evolution of the spin state of Ajisai in accordance with the empirical models is shown in Figure \ref{fig:ajisai_empirical}. The equations to calculate the spin-axis vector and period are given in Appendix \ref{appendix:empirical_model_cals}.

\section{Instrumentation and Observations}

The SuperWASP (Super Wide Angle Search for Planets) project \citep{Pollacco2006} was a highly successful ground-based exoplanet survey. Operating from 2004 until 2018, it discovered and confirmed nearly 200 transiting Hot Jupiter class planets. The survey operated two custom survey telescopes (SuperWASP North on La Palma and SuperWASP South in South Africa) that used an array of consumer telephoto lenses (the Canon EF 200mm f/1.8L USM) with scientific CCD cameras to provide an instantaneous field of view of nearly 500 deg$^2$ for each telescope.

\begin{figure}
	\centering
	\includegraphics[width=0.48\textwidth]{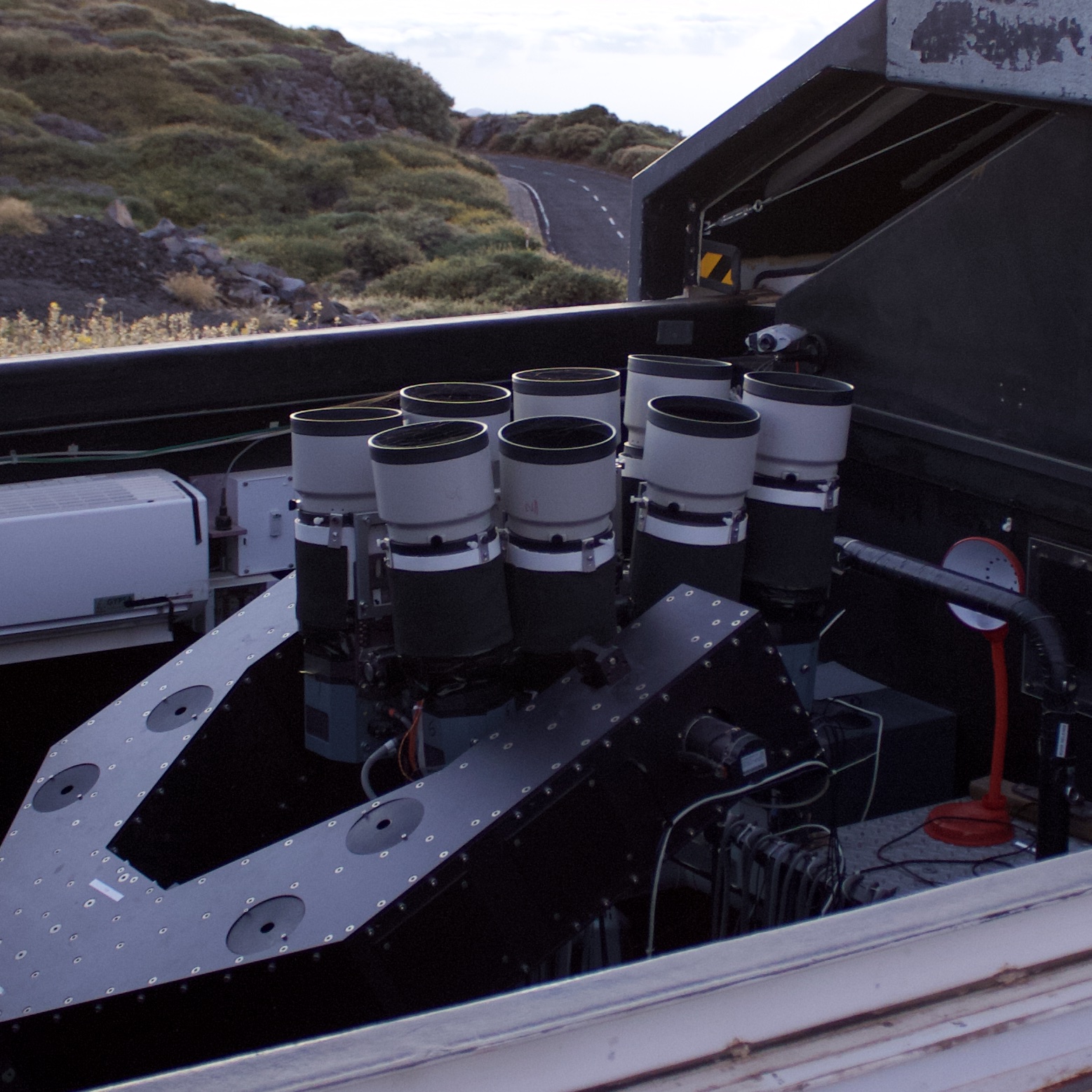}
	\caption{SuperWASP North at the Roque de los Muchachos Observatory in August 2019.}\label{fig:swasp-north}
\end{figure}

Between 2018 -- 2022, SuperWASP North (hereafter just SWASP) was used for several observational Space Domain Awareness (SDA) research projects. A picture of SWASP during this period is shown in Figure \ref{fig:swasp-north}.

The SuperWASP project was searching for transit signals that occur on timescales of minutes to hours; as such, it was not designed to support the sub-second timestamp accuracy required when observing fast-moving LEO satellites. In particular, variable timing delays between the Telescope Control Server (TCS) and the individual Data Acquisition Servers (DASes) for each camera meant that the individual mosaic sub-exposure start times could differ by more than a second. A further problem arose from the dated Scientific Linux 3 operating system that was not compatible with modern scientific software libraries.

\begin{figure*}
	\centering
	\includegraphics[height=70mm]{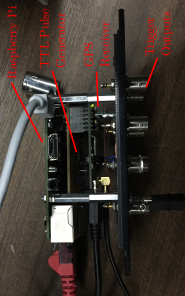}
	\includegraphics[height=70mm, trim=0 4mm 7mm 5mm]{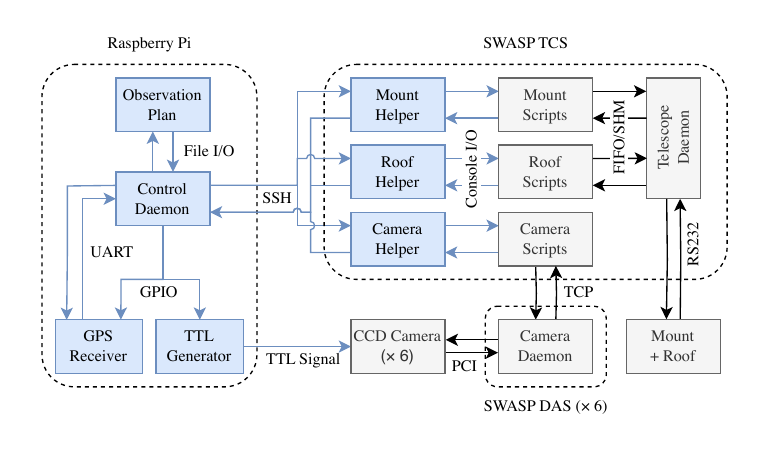}
	\caption{\textbf{Left:~}The custom \textit{Raspberry Pi} control system. \textbf{Right:~}A high-level schematic of the modified system operations, with new components shaded in blue and the original SWASP control system shaded grey. See text for more details.}\label{fig:gpstrigger}
\end{figure*}

These issues were resolved by introducing a \textit{Raspberry Pi} single-board computer to control observations. A custom control daemon (implemented using Python) parsed an observation plan file (containing a list of positions and times) and issued shell commands via SSH to custom helper programs (implemented using C) on the TCS that then called the original SWASP control system scripts to point the mount and prepare exposures. Custom circuitry, extending the principles of an earlier instrument, \textit{Puoko-nui} \citep{Chote2014}, generated a GPS-timestamped TTL pulse to simultaneously trigger the cameras using their external trigger input. The timestamps and associated exposure IDs were saved to the observing plan file where they could be reconciled during subsequent processing. Figure \ref{fig:gpstrigger} provides a high-level overview of this modified system.

By 2018, only six of the original eight CCD cameras were operational. These were realigned using the per-camera adjustments available on the mount, with four cameras forming a square $16^\circ\times16^\circ$ field of view and the remaining two cameras overlapping the central field to provide redundancy and to enable cross-validation between cameras.

\subsection{LEO Survey}
The modified SWASP undertook a survey of LEO resident space objects (RSOs) between July -- November 2019, and captured in total 1826 passes of 774 individual RSOs. 

The Observatorio del Roque de Los Muchachos on La Palma is an excellent dark-sky site (2349\,m altitude; V\textsubscript{sky} $\sim$ 22\,mag/arcsec\textsuperscript{2}), but the large angular size of the SuperWASP pixels meant that the effective sky background during dark conditions was V\textsubscript{sky} $\sim$ 16.5\,mag. This limited its usefulness to observing stars brighter than V $\sim$ 15 \citep{Pollacco2006}, but was well matched to the population of TLE-catalogued LEO RSOs, which are generally V $\lesssim$ 12 mag.

The path of a target RSO across the sky was divided into a set of precisely timed sidereal fields that capture the target shortly after it entered the field of view, and completed shortly before it left. The exposure time for each field was determined by the time it took the RSO to cross the field, which was typically a few seconds (and limited to a maximum of 45 seconds to avoid saturation of background stars). 

High time resolution light curves were extracted by measuring the change in surface brightness along the length of the streak; the effective time resolution was set by the time it takes the target to move by a point-spread-function width across the CCD; of order tens of milliseconds at typical LEO altitudes for SuperWASP's 13 arcsecond pixel size.

Each exposure was separated by a 15 second delay, during which the CCDs were read out and the mount repositioned to the next field. The same sidereal pointings were observed for a second time immediately after each pass (generally within $5 - 10$ minutes, minimising any potential changes in sky transparency) to provide reference images of the sky behind the target streaks. Difference images were created using \textsc{hotpants} \citep{Becker2015} to align, resample, and subtract the images. Pixels associated with bright sources in the reference image were masked in the difference image to reduce false signals introduced by residual difference artifacts.

\begin{figure*}
	\centering
	\includegraphics{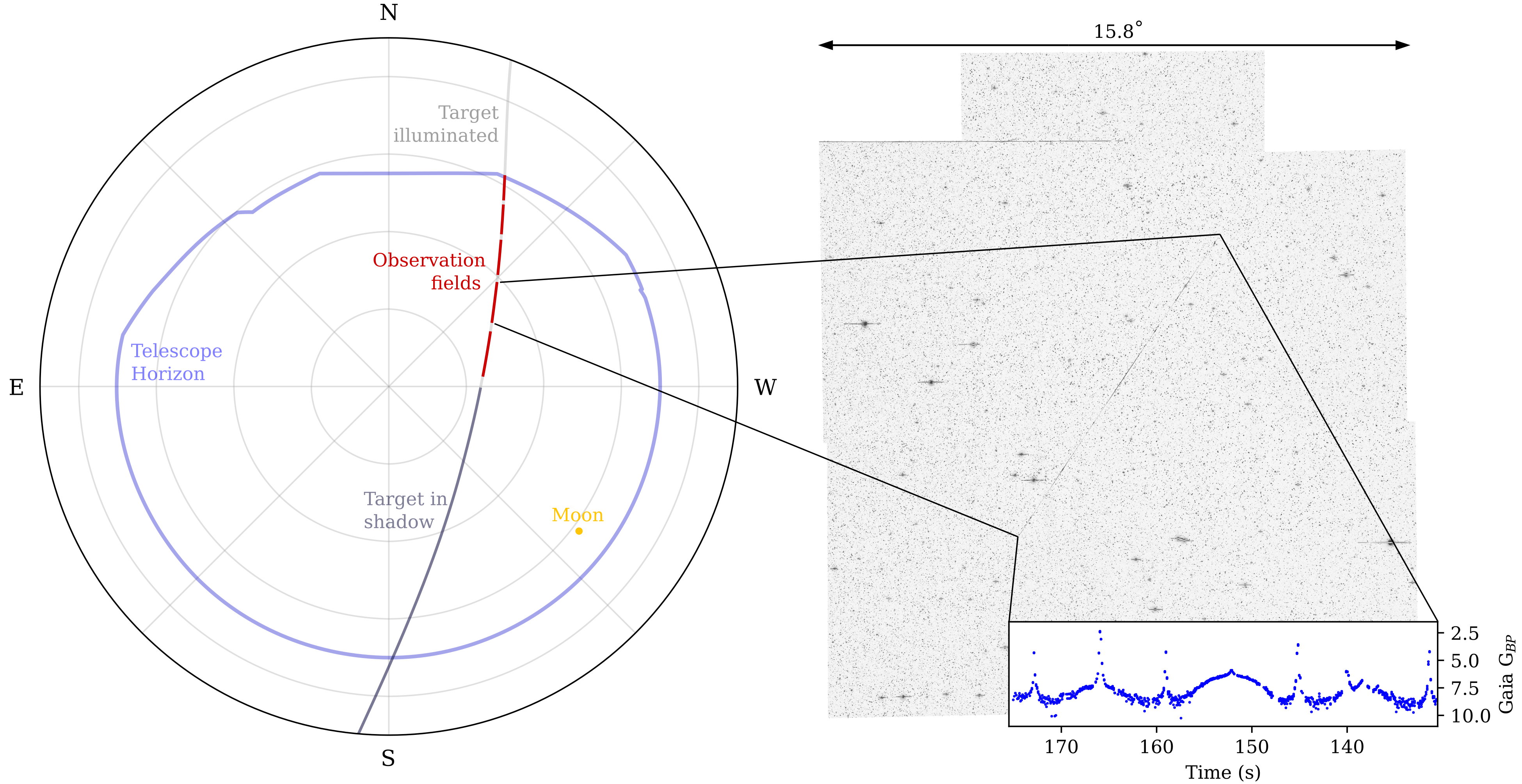}	
	\caption{A schematic illustration of the LEO observation procedure using SuperWASP. \textbf{Left:} The pass of the target across the sky is divided into a series of fields that account for the visibility limits of the telescope, shadow cone of the Earth, separation from the Moon, and dead time while the telescope slews and cameras read out. \textbf{Right:} The footprint of a single pointing (shown as a mosaic of the 6 cameras) captures a tumbling RSO as it streaks across the field. Photometry is extracted from the streak with an effective time resolution set by the speed of the target across the CCDs. Figure reproduced from \protect\cite{Chote2019}.}\label{fig:swasp_plan}
\end{figure*}

Figure \ref{fig:swasp_plan} illustrates the observation procedure, which was described in detail, along with the data reduction pipeline,  in \cite{Chote2019}.

\subsection{Ajisai Observations}

\begin{figure*}
    \centering
    \includegraphics{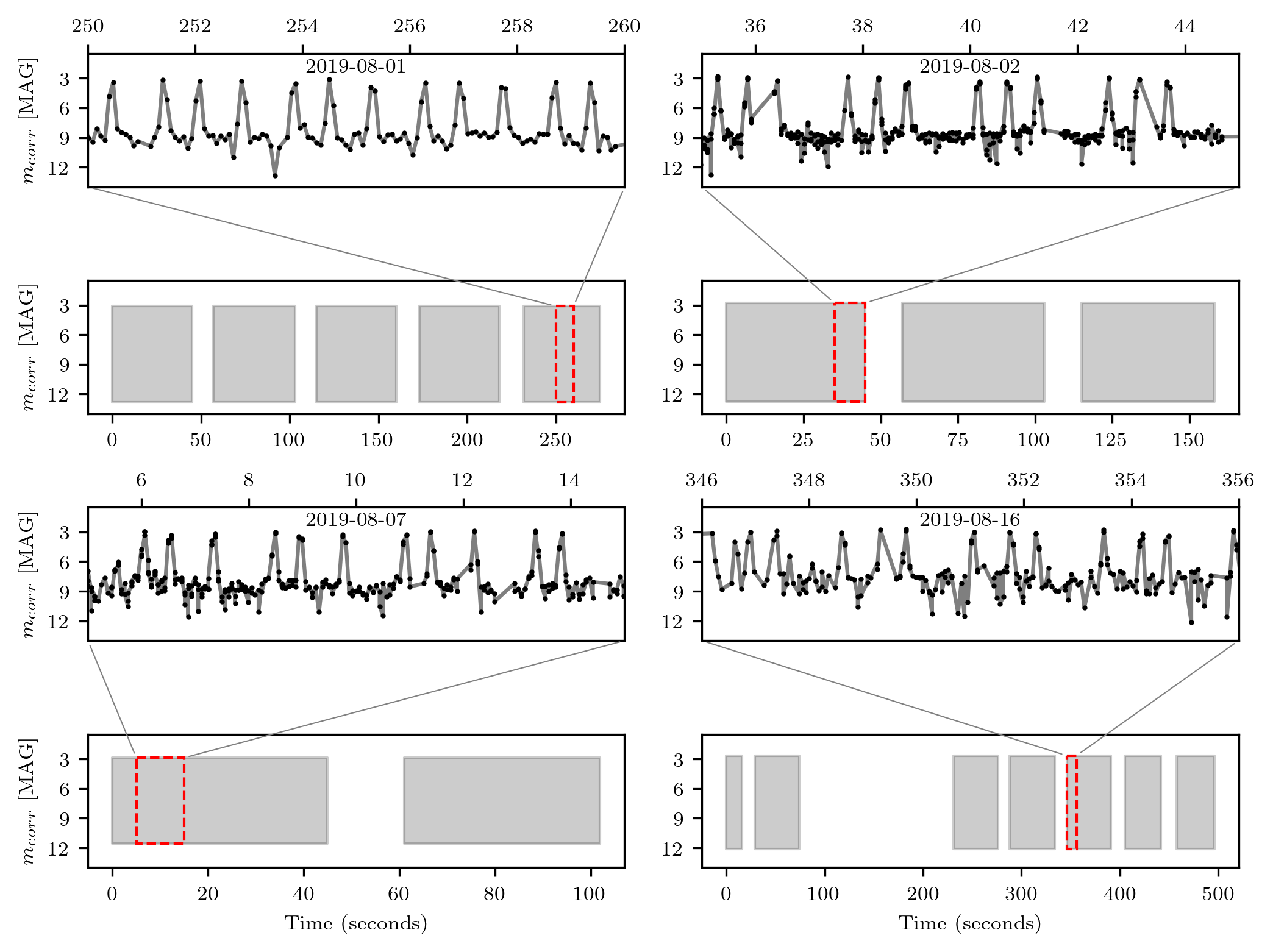}
    \caption{Summary of the SuperWASP observations of the Ajisai satellite. The upper panel of each observation shows a small section of the light curve. Because the spin period $\sim 2.5\,$s is significantly shorter than the observation lengths (tens to hundreds of seconds), a plot of the full light curves are not able to resolve the individual glints; for visual simplicity we instead represent the full light curves in the lower panels as shaded regions that indicate the time ranges of each field and the inter-field gaps. The dashed red rectangular boxes display the bounds of the zoomed in region. The triplet features occurring due to mirror reflections are clearly demonstrated within the zoomed in regions. All brightness measurements are range-corrected to 1000$\,km$.}\label{fig:ajisai_overview}
\end{figure*}

Ajisai was observed 4 times between the nights of the 2019~August~1~--~16. Figure \ref{fig:ajisai_overview} and Table \ref{tab:obs} provide a summary of these data. 

\begin{figure}
	\centering
	\includegraphics{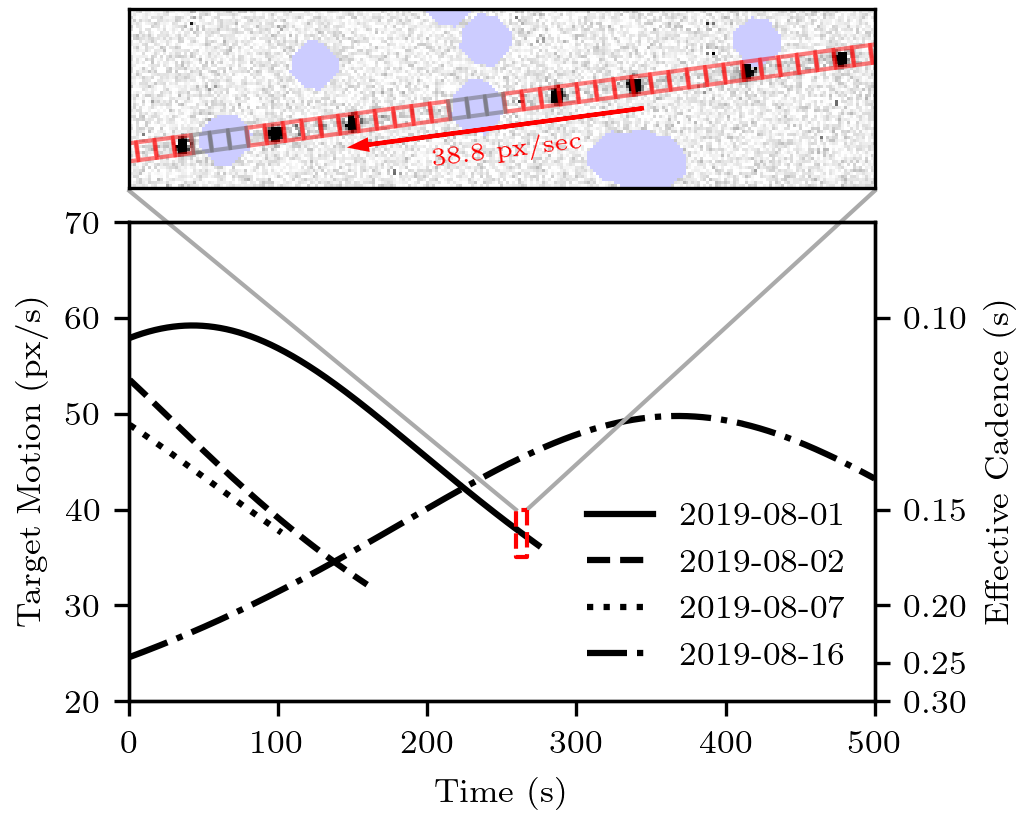}
	\caption{Photometry was extracted from the target streaks using a fixed aperture size of $6\times6$ pixels, chosen to avoid over-sampling the spatial resolution of the image. \textbf{Bottom:} The target motion and resulting extraction cadence are plotted for the four Ajisai observations. \textbf{Top:} An example of a difference image segment with apertures visualised. Ajisai is visible as a faint continuous streak interspersed with point-source-like glints. Masked pixels containing difference-image artefacts are tinted blue, and apertures that intersect these regions are excluded.}\label{fig:extractioncadence}
\end{figure}

The rate of motion of Ajisai compared to the background stars changes during each pass, producing a time resolution that is continuously changing. Figure~\ref{fig:extractioncadence} illustrates that the typical time resolution is $100-150\,$ms: an order of magnitude larger than the $\sim 10\,$ms wide glint features \citep{CALATRONI2025}. The photometric map methodology described in Section \ref{sec:methodology} is sensitive primarily to the spacing between glints, so it is not necessary to resolve the individual glints.

\begin{table}
    \caption{Summary of our 2019 SuperWASP observations of Ajisai. $D_{\text{obs}}$, $N_{\text{data}}$ and $N_{\text{fields}}$ refer to the number of days since launch ($D_{\text{launch}} = 46654.86$ [MJD]), the number of observed datapoints and the number of sidereal fields the satellite was observed in respectively.}
    \centering
    \resizebox{\linewidth}{!}{%
    \begin{tabular}{|c|c|c|c|c|}
    \hline
        $T_{\text{start}}$ [UTC] & $T_{\text{end}}$ [UTC] & $D_{\text{obs}}$ [Days] & $N_{\text{data}}$ & $N_{\text{fields}}$ \\ \hline
         2019-08-02 3:25:03 & 2019-08-02 3:29:38 & 12042.28  & 6787 & 5  \\ \hline
         2019-08-03 2:33:25 & 2019-08-03 2:35:18 & 12043.25   & 3665 & 3 \\ \hline
         2019-08-08 2:06:06 & 2019-08-08 2:07:55 & 12048.23  & 1597 & 2  \\ \hline
         2019-08-16 23:56:23 & 2019-08-17 00:05:42 & 12057.14  & 6224 & 7  \\ \hline
    \end{tabular}}

    \label{tab:obs}
\end{table}

\section{Methodology}\label{sec:methodology}

The procedure for extracting the spin-axis and period of Ajisai from our photometric observations can be broadly broken down into the following steps:
\begin{enumerate}
    \item Defining a function that calculates (from an input Two Line Element ephemeris) the Phase Angle Bisector (PAB) vector in the ICRF reference frame (see section \ref{sec:transforms/vectors}) as a function of input time.
    \item Defining a function that transforms the PAB unit vectors, for a chosen set of rotation parameters (spin-pole coordinates, period and a rotational phase offset), into the RSO body fixed frame.
    \item Using these two functions to map each measurement in the light curve, for a chosen set of rotation parameters, to a latitude and longitude on the surface of Ajisai.
    \item Comparing the resulting two dimensional glint positions against the known mirror positions and using a MCMC (Markov Chain Monte Carlo) search to find the parameters (and associated uncertainties) that produce the best alignment.
\end{enumerate}

The result of this procedure is a two dimensional brightness map that visualises the surface reflectivity of Ajisai. The position of each mirror (strictly speaking, the orientation of its normal vector) is clearly visible by a localised spike in brightness centered on its position.

\subsection{Reference Frame Transformation and Vector Calculations}
\label{sec:transforms/vectors}
\begin{figure}
    \centering
    \includegraphics[width=0.5\textwidth]{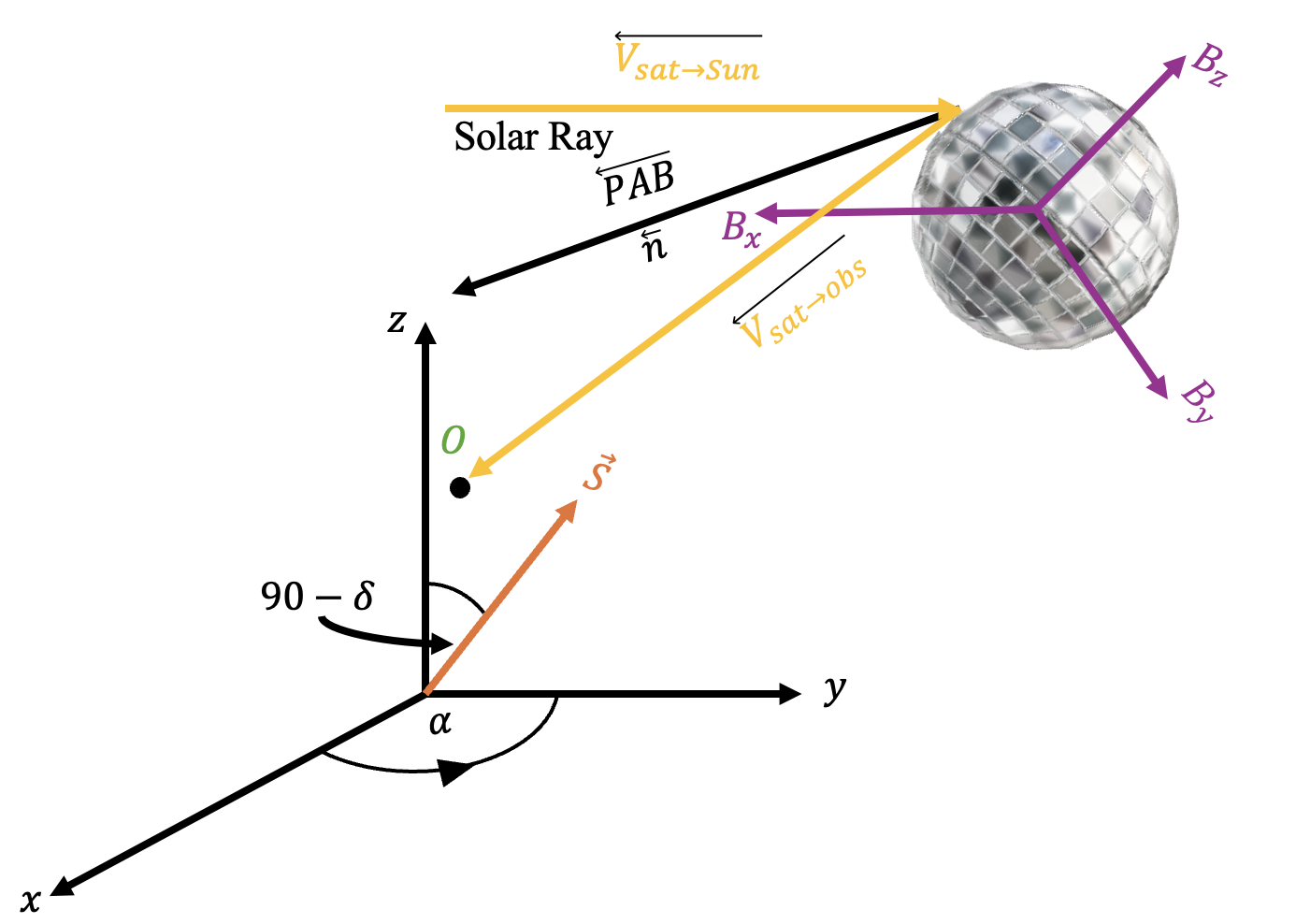}
    \caption{Diagram illustrating the defined angles, $\alpha$ and $\delta$ of the spin-axis vector, $\vec{S}$ with respect to the axes of the ICRF. The PAB is also defined with respect to the vector directions toward the Sun and observer, O, from the satellite with body frame axes, B.}
    \label{fig:vector_angles}
\end{figure}

The PAB, p, is defined as the unit vector halfway between the satellite-to-Sun ($V_{\text{sat$\rightarrow$ Sun}}$) and satellite-to-observer directions ($V_{\text{sat$\rightarrow$obs}}$):
\begin{equation}
    p = \frac{ V_{\text{sat$\rightarrow$obs}}(t) + V_{\text{sat$\rightarrow$ Sun}}(t)}{\| V_{\text{sat$\rightarrow$obs}}(t) + V_{\text{sat$\rightarrow$ Sun}}(t)\|}
    \label{eq:pab_calc}
\end{equation}
It is a useful quantity in determining whether a reflection has occurred off a surface as the specular condition holds true that a reflection will occur when the PAB vector and surface normal, $\vec{n}$ to a flat facet nearly coincide \citep{2013amos.confE..34H}. The geometry of the PAB and angles related to the frame transform are illustrated in Figure \ref{fig:vector_angles}.

In order to compare our observations with the model, we must transform the PAB vectors from the ICRF into the body-fixed frame. This transform depends on the spin state of the satellite such that finding the optimum spin parameters for this transformation will determine how well we map the surface reflectivity of the satellite in the body fixed frame of the satellite.

\begin{figure}
    \centering
    \includegraphics[width = 0.5\textwidth]{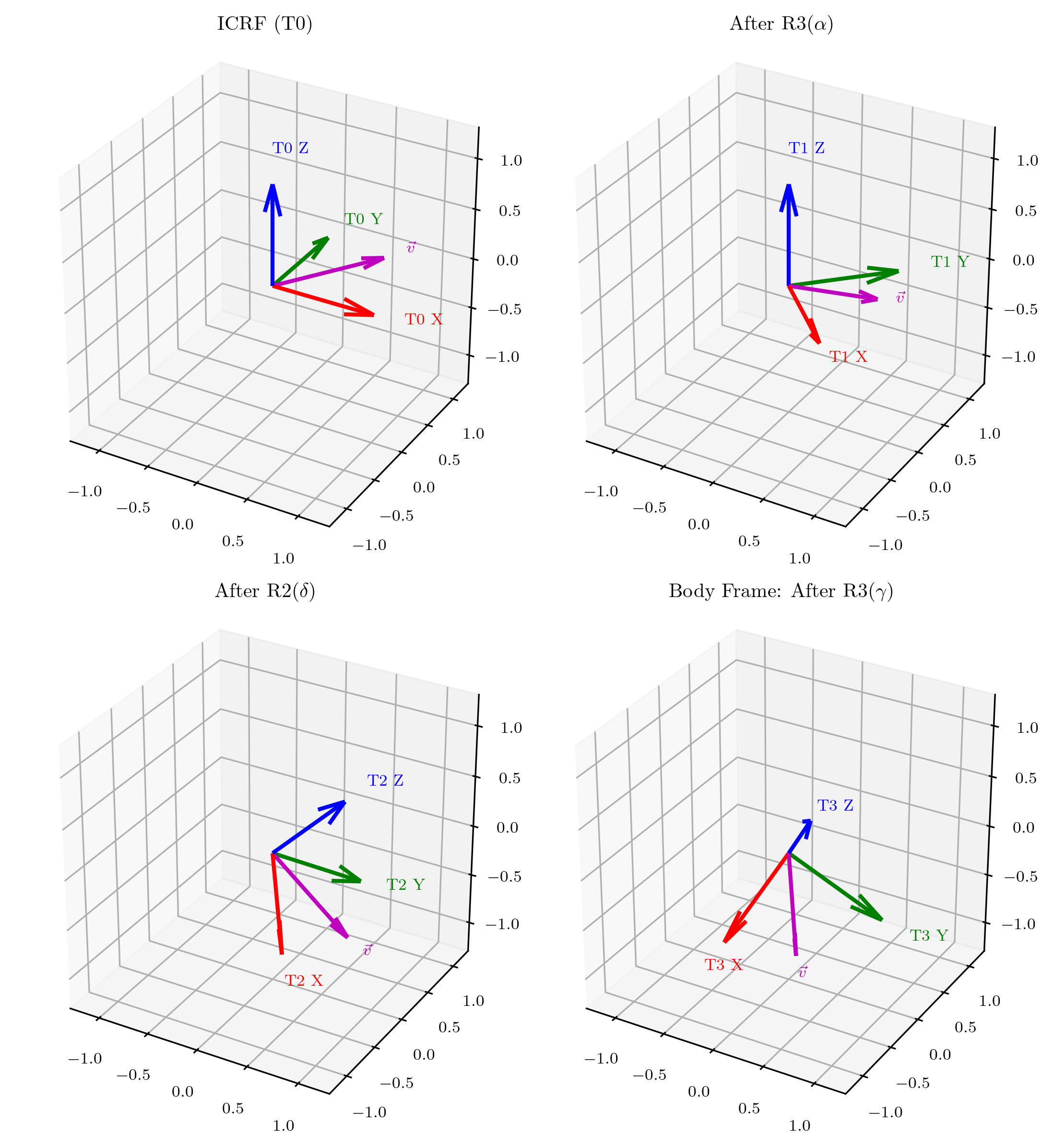}
    \caption{Illustration of the rotation sequence to transform a vector, $\vec{v}$ from the ICRF to body frame. In this example, both the axes (T0) and vector are transformed by $\alpha$,$\delta$ and $\gamma$ of 45$^\circ$.}
    \label{fig:vector_transform}
\end{figure}

We are able to transform the satellite-to-Sun and satellite-to-observer vectors from the ICRF frame to the body frame by using a transformation matrix, T, which follows standard right hand rotation matrices around the z ($R_{3}$) and y ($R_{2}$) axes respectively \citep[see e.g.,][]{1967AmJPh..35.1097A,1980clme.book.....G,Seeber+2003}.
\begin{equation}
R_{3}(\gamma) =
\begin{bmatrix}
\cos{\gamma} & \sin{\gamma} & 0 \\
-\sin{\gamma} & \cos{\gamma} & 0 \\
0 & 0 & 1
\label{eq:R3-gamma}
\end{bmatrix}
\end{equation}\\
\begin{equation}
R_{2}(\delta) = 
\begin{bmatrix}
\cos{(\frac{\pi}{2} - \delta)} & 0 & \sin{(\frac{\pi}{2} - \delta)} \\
0 & 1 & 0 \\
\sin{(\frac{\pi}{2} - \delta)} & 0 & \cos{(\frac{\pi}{2} - \delta)}
\label{eq:R2}
\end{bmatrix}
\end{equation}
\begin{equation}
R_{3}(\alpha) =
\begin{bmatrix}
\cos{\alpha} & \sin{\alpha} & 0 \\
-\sin{\alpha} & \cos{\alpha} & 0 \\
0 & 0 & 1
\label{eq:R3-alpha}
\end{bmatrix}
\end{equation}
\begin{equation}
 T = R_3(\gamma)\cdot R_2(\delta) \cdot R_3(\alpha)
\label{eq:transformation_matrix}
\end{equation} 
where $\gamma$ is the spin-angle, and $\alpha$, $\delta$ are the right ascension and declination of the spin-pole in the ICRF respectively. The first rotation, $R_3(\alpha)$ aligns the x-axis with the spin-pole's projection onto the equatorial plane. The second rotation tilts the new coordinate system such that the new z-axis points towards the spin-pole. Finally, the third rotation sets the orientation of the satellite about its spin-axis at a given time-step, essentially defining the rotation around the newly aligned z-axis. Figure \ref{fig:vector_transform} illustrates the sequence of rotations from ICRF to body frame.
\begin{figure*}    
    \centering
    \includegraphics{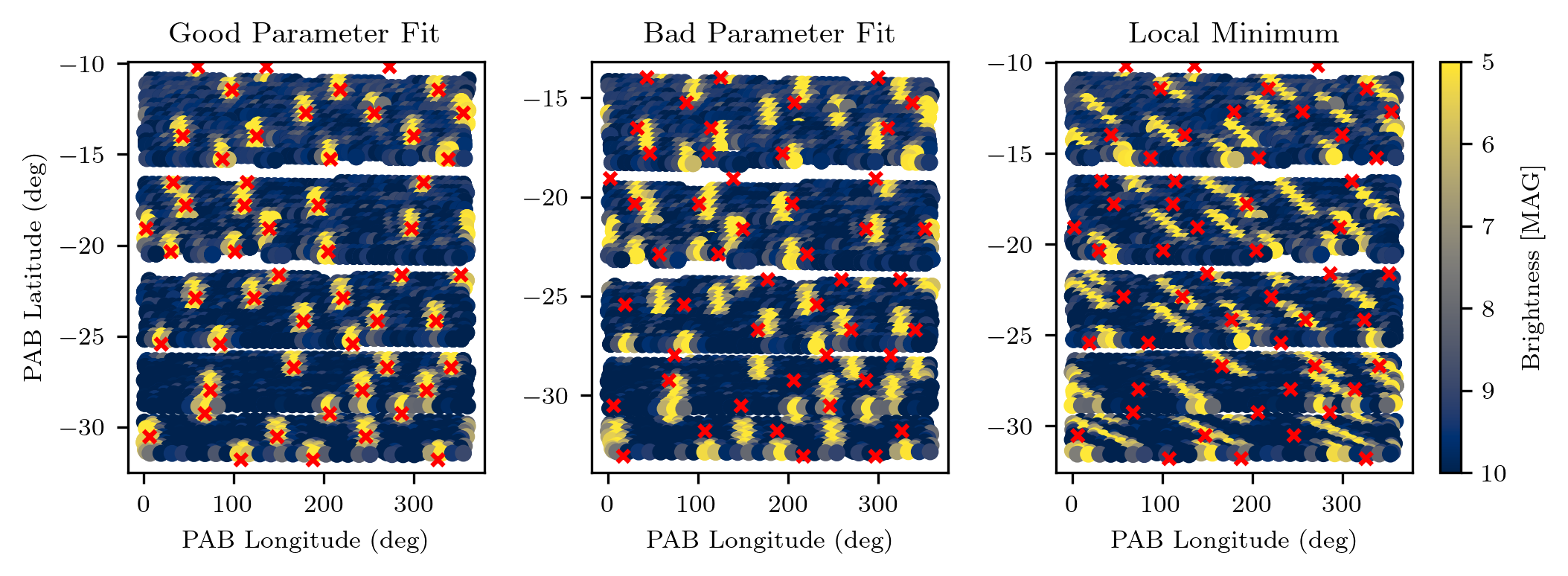}
    \caption{Examples of brightness maps from the 2019-08-01 observation with three different parameter sets are compared to the \protect\cite{2017AdSpR..60.1389K} mirror positions (red crosses). \textbf{Left:} The best-fit parameters transform all of the photometric glints onto an expected mirror position. \textbf{Centre:} A poor fit is produced by offsetting the $\delta$ of the best fitted solution by 5 degrees. \textbf{Right:} Local minima are found in areas of the parameter space that project some of the photometric glints onto an expected mirror position, but project others in the gaps between mirrors.}
    \label{fig:good_bad}
\end{figure*}

The satellite-Sun, satellite-observer and consequently the PAB vectors were calculated (see Equation \ref{eq:pab_calc}) in Python using \textsc{Skyfield} \citep{skyfield}, which effectively propagates a TLE into an orbital position from an observer location with respect to the desired reference frame, in this case, the ICRF. The vectors were calculated from the TLEs at a 2 millisecond cadence between the start and end of each observation window. As a performance optimisation, these calculations were performed once and fitted with an interpolating spline, which was subsequently evaluated at the observation times.

The PAB unit vectors were then transformed into the body-fixed frame through the application of the transformation matrix, T, shown by equation \ref{eq:transformation_matrix} and the latitudinal ($\psi$) and longitudinal ($\theta$) components were extracted from the vector form via equations \ref{eq:lat} and \ref{eq:lon}, where $v$ is the unit PAB vector in the body frame.

\begin{equation}
    \psi = \arcsin{v_z}
    \label{eq:lat}
\end{equation}
\begin{equation}
    \theta = \arctan2(v_y,v_x) + \pi
    \label{eq:lon}
\end{equation}

This approach allows the flux measurements to be associated with a coordinate pair and thus we can create a brightness map of the Ajisai surface.

\subsection{Interpreting Brightness Maps}

Brightness maps represent a novel visualisation of the surface reflectivity of the satellite as a function of the derived spin-state solution. Each map is plotted in latitude–longitude body-frame PAB coordinates, where each point is associated with a measured brightness value (magnitude) indicated by the accompanying colour scale.
Mirror positions from \cite{2017AdSpR..60.1389K} are overlaid as crosses:
\begin{itemize}
    \item Black crosses define mirrors that were not observed.
    \item Red crosses define mirrors that were observed.
\end{itemize}

Better spin-state solutions produce glint patterns that align more closely with these reference mirror positions (see Figure \ref{fig:good_bad}).
The effect of sub-optimal solutions is also illustrated. A solution trapped in a local minimum may still produce partial alignment—some glints coincide with reference mirrors—yet most remain offset. Such a solution scores better than a poor (misaligned) fit but remains inferior to a fully aligned, optimal spin-state solution.

\subsection{Fitting}
The lat-lon coordinates can be compared to the modelled mirror points from \cite{2017AdSpR..60.1389K} in terms of spherical distance, which accounts for the periodic boundaries imposed by the longitudinal component.

\begin{table*}
    \caption{Table of priors corresponding to the observation on the night of 2019-08-07 (see Figure \ref{fig:corner_2x2}). The prior distributions are given in standard probability theory notation form \citep{gelman2013bayesian}. The initial phase offset distribution was set to search the full bounds but was then iterated upon after initial exploration as we have no prior knowledge of the phase offset. The convention here is assuming the spin-pole of Ajisai lies towards the north celestial pole and therefore the spin direction is counter-clockwise -- hence the negative rotation period.}
    \centering
    \begin{tabular}{|c|c|c|c|c|c|}
        \hline
        Parameter & Description & Prior Distribution & Support & Units & Justification \\
        \hline
        $\alpha$ & Spin-pole RA & $\mathcal{N}(\alpha_{\text{emp}},\, 14^2)$ & $[-140, -75]$ & $^\circ$ & Empirical RA + residuals \\
        \hline
        $\delta$ & Spin-pole Dec & $\mathcal{N}(\delta_{\text{emp}},\, 0.5^2)$ & $[80, 90]$ & $^\circ$ & Empirical Dec + residuals \\
        \hline
        $P$ & Rotation Period & $\mathcal{N}(P_{\text{emp}},\, (0.000412)^2)$ & $[-2.4355, -2.4322]$ & $s$ & Empirical period + RMS \\
        \hline
        $\phi$ & Phase Offset & $\mathcal{N}(180,\, 10^2)$ & $[0, 360]$ & $^\circ$ & Full phase range centered \\
        \hline
    \end{tabular}
    
    \label{tab:priors}
\end{table*}

Four parameters are required to transform the PAB vectors from the ICRF to the body-fixed frame: the spin-pole right ascension ($\alpha$), declination ($\delta$), rotation period ($P$), and phase offset ($\phi$). The rotation period determines the angular velocity applied at each time step (see Equation \ref{eq:R3-gamma}), while the phase offset accounts for the satellite’s initial orientation at the start of the observation.

To assess how well a given parameter set localises glint flux around known mirror orientations, we define a scoring metric (Equation \ref{eq:fitting_metric}) that minimises the flux-weighted squared angular distance (see Appendix \ref{appendix:sep_calc} for details on the angular separation calculation) between interpolated points and the closest mirror point. This penalises solutions that produce significant flux away from mirror locations. Well-aligned fits produce sharp, localized spikes, whereas poor fits smear them out. This behaviour could, in principle, be used to constrain the positions of unknown mirrors without relying on a priori knowledge.
\begin{equation}
\begin{aligned}
\min_{\mathbf{(\alpha,\,\delta,\,P,\,\phi)}}\sum_{i=1}^{n} f_i \cdot \min_{j} \left\{ \text{sep}(\mathbf{x}_i, \mathbf{y}_j) \right\}^2
\label{eq:fitting_metric}
\end{aligned}
\end{equation}

\text{Subject to:}
\begin{align}
\mathbf{x}_i &= (\text{ilon}_i, \text{ilat}_i) \quad \forall i \in \{1, \ldots, n\} \\
\mathbf{y}_j &= (\text{K-Lon}_j, \text{K-Lat}_j) \quad \forall j \in \{1, \ldots, m\} \label{eq:subect_koshkin}
\end{align}

\text{Where:}
\begin{itemize}
\item $n$ is the number of interpolated points ($ilon$, $ilat$)
\item $m$ is the number of mirror points in the dataset 
(\text{K-Lon}, \text{K-Lat})
\item $f_i$ is the flux value for the $i$-th interpolated point
\item $\text{sep}(\mathbf{x}_i, \mathbf{y}_j)$ is the angular separation between coordinates $\mathbf{x}_i$ and $\mathbf{y}_j$ in degrees
\item $\mathbf{x}_i$ represents the coordinates of the $i$-th interpolated point
\item $\mathbf{y}_j$ represents the coordinates of the $j$-th point in the full dataset of mirror points
\end{itemize}

We implemented an MCMC sampler using \textsc{emcee} \citep{2013PASP..125..306F} to explore the posterior distribution of these parameters. An example set of prior distributions are listed in Table \ref{tab:priors}, and the empirical estimates with uncertainties reflecting residuals and RMS errors from \cite{2016AdSpR..57..983K} are taken into consideration for the specific search band for that observation. Specifically, the empirical uncertainties are approximately $\alpha_{\sigma} \sim 14^\circ$, $\delta_{\sigma} \sim 0.5^\circ$, and $P_{\sigma} \sim 0.412 \times 10^{-3}$s. An initial exploration was performed using wider search bands for each parameter, but the uncertainties of the fit were characterised across search bands similar to those referenced in Table \ref{tab:priors}.

\begin{figure*}
  \centering
  \setlength{\tabcolsep}{0.5pt} 
  \renewcommand{\arraystretch}{0} 
  \begin{tabular}{cc}
    \includegraphics{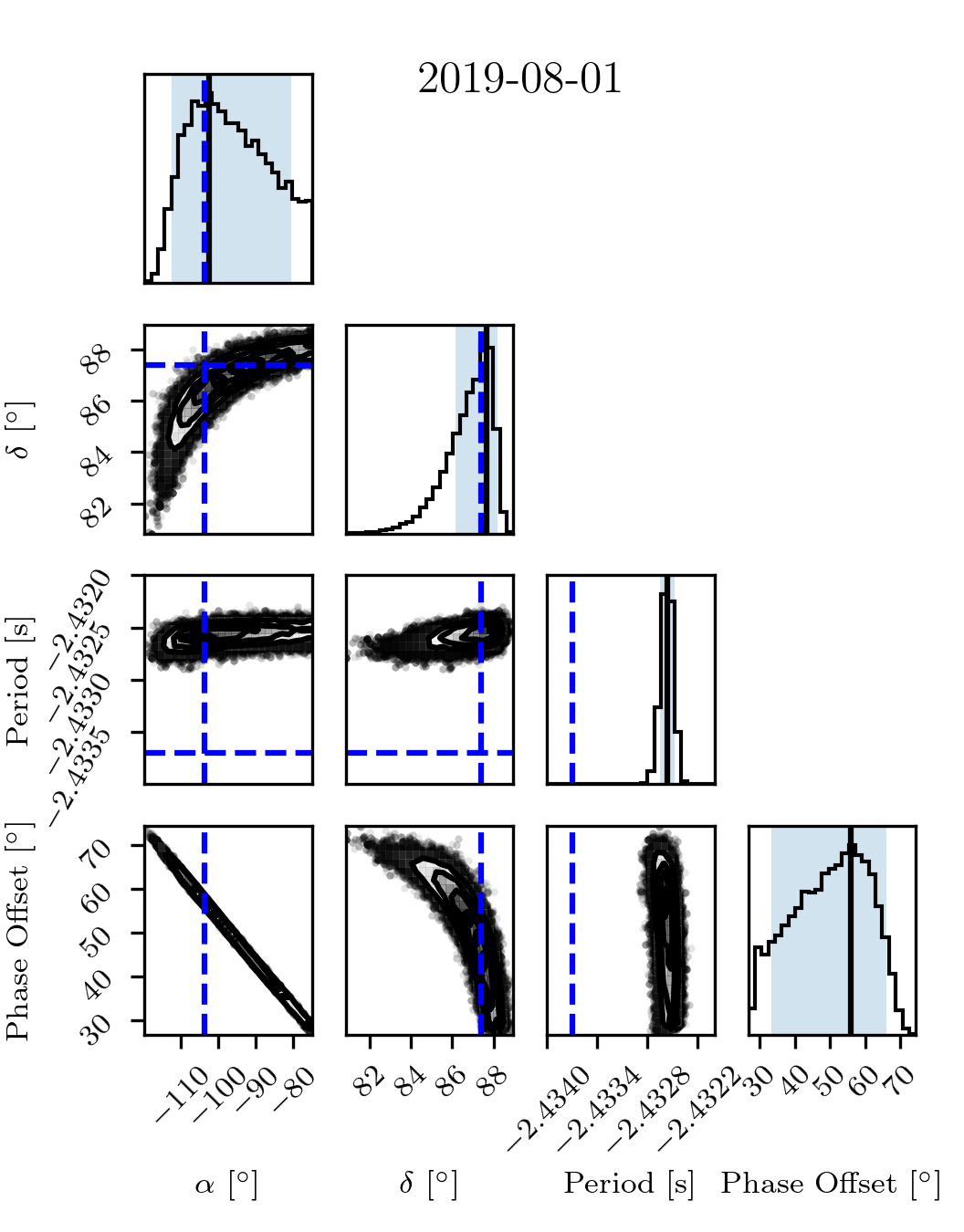} &
    \includegraphics{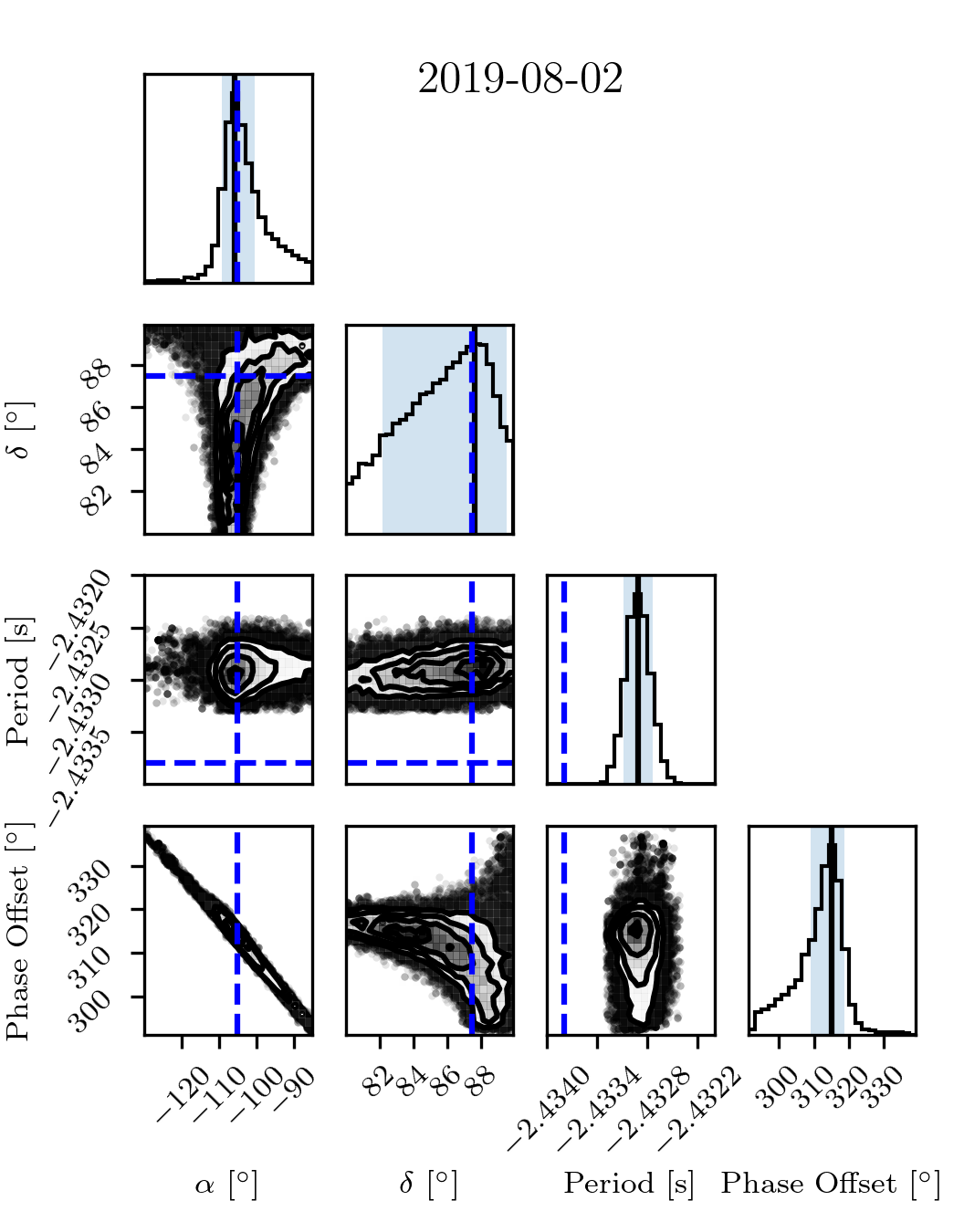} \\[1pt]
    \includegraphics{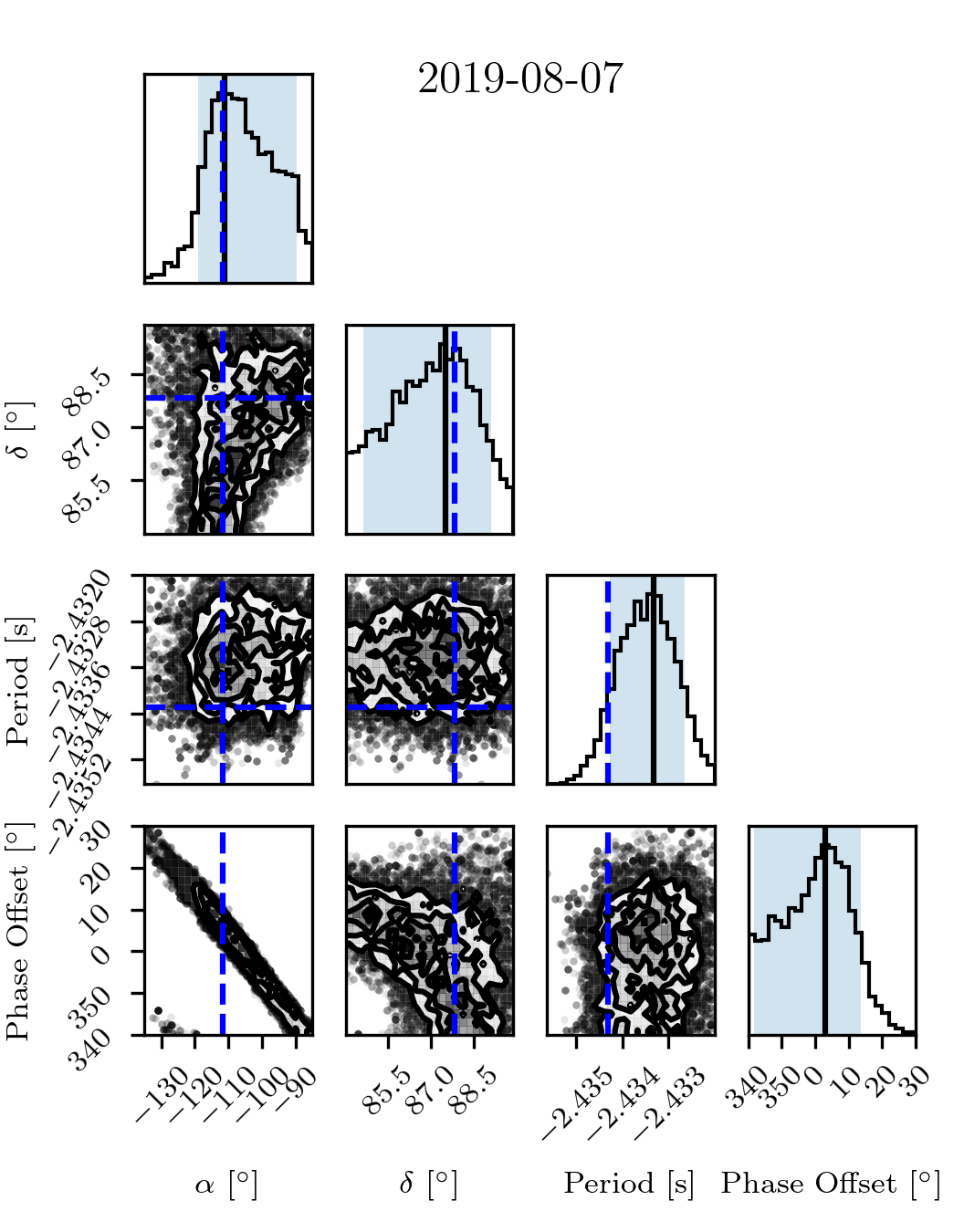} &
    \includegraphics{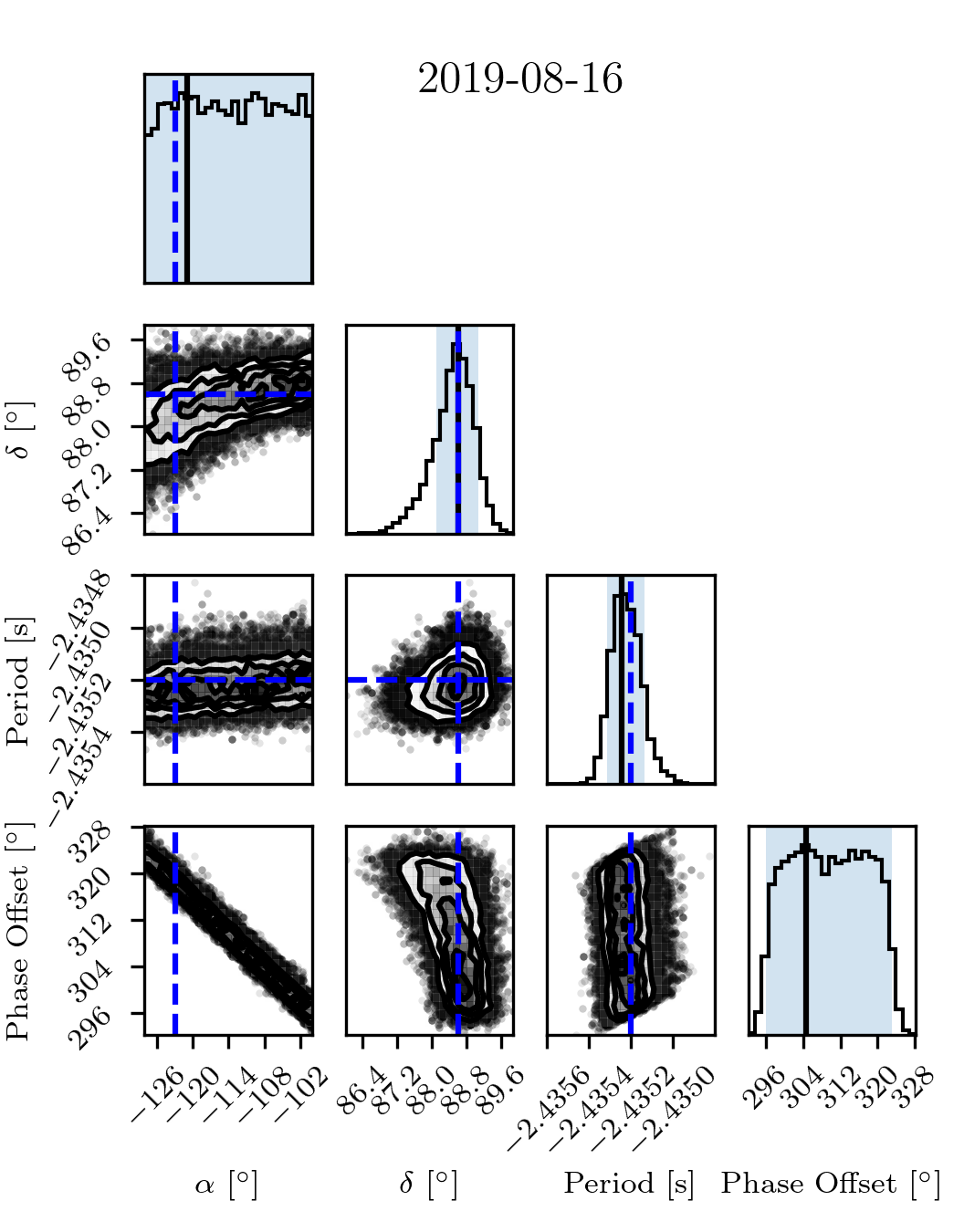}
  \end{tabular}
    \caption{Corner plots from the MCMC runs across all four observations. The blue dashed lines indicate the empirical value estimate from \citep{2016AdSpR..57..983K}, with the black lines and blue shaded regions representing the best estimate and error bounds from the mode and FWHM of the sample distributions. In some instances, the empirical estimate (e.g. rotational period) do not pass through a minimum, this may be indicative of variations for example, seasonal illumination changes. Contours show 2-D highest-posterior-density credible regions from the MCMC samples; darker shading indicates higher posterior density.}
    \label{fig:corner_2x2}
\end{figure*}

Each MCMC production run consisted of 5000 steps, with the first 1000 steps discarded as burn-in. The best-fit parameters and their uncertainties were estimated using the mode and full width half maximum (FWHM) of the posterior distributions. The corresponding corner plots from each run are shown in Figure \ref{fig:corner_2x2} and are generated using \textsc{corner} \citep{corner}.

\begin{figure*}
    \centering\includegraphics{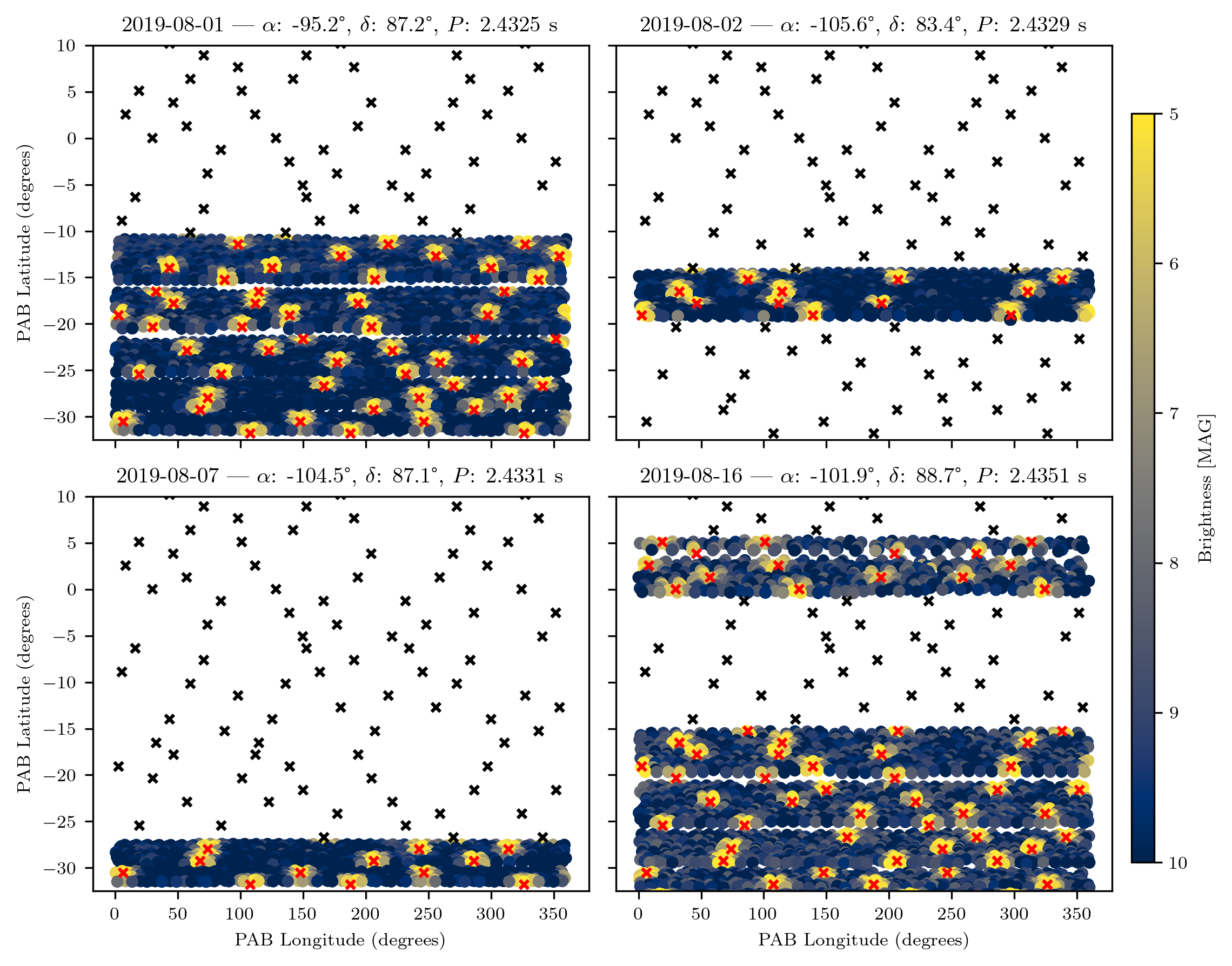}
    \caption{Brightness maps generated using the MCMC parameter sets (shown in the subplot titles) with the maximum log probability parameter sets for each observation, which are given above each brightness map. Red crosses denote the reference mirror points, overlapping with coloured points that indicate interpolated body-fixed lat-lon coordinates and associated magnitude values. Mirror position data is from \protect\cite{2017AdSpR..60.1389K}, used with permission.}
    \label{fig:glint_map_results}
\end{figure*}
\section{Results}

\begin{table*}
    \caption{Best estimates of parameters and their uncertainties derived from post burn-in MCMC samples. The estimate and asymmetric errors are derived from taking the mode and FWHM of the sample distributions. The alpha ($\alpha$) and delta ($\delta$) of the empirical spin-pole (ICRF frame) and period calculated at time of observation are given and we note the spin-axis, $\hat\Psi_{\text{emp}} (\alpha_{\text{emp}},\delta_{\text{emp}})$ is inverted as this forms a right hand coordinate system such to complete the reference frame transforms, consistent with \protect\cite{2017AdSpR..60.1389K}. In this convention, the spin-pole of Ajisai is pointing toward the north celestial pole and whilst the absolute value for the rotation period is given, the spin-direction would be counter-clockwise.}
    \centering
    \begin{tabular}{|c|c|c|c|c|c|c|}
    \hline
        Night Observed & $\alpha$ [$^\circ$] & $\delta$ [$^\circ$] & $|P|$ [s] & $\alpha_{\text{emp}}$ [$^\circ$] & $\delta_{\text{emp}}$ [$^\circ$] & $P_{\text{emp}}$ [s] \\ \hline
        2019-08-01 & $-102.40\,^{+21.81}_{-10.01}$ & $87.65\,^{+0.53}_{-1.49}$ & $2.4326\,^{+0.00009}_{-0.00009}$ & $-103.67$ & $87.39$ & 2.4337 \\ \hline
        2019-08-02 & $-105.89\,^{+5.40}_{-3.42}$ & $87.59\,^{+1.93}_{-5.43}$ & $2.4329\,^{+0.00017}_{-0.00017}$ & $-105.0$ & $87.46$ & 2.4338 \\ \hline
        2019-08-07 & $-111.19\,^{+21.39}_{-7.77}$ & $87.49\,^{+1.61}_{-2.87}$ & $2.4333\,^{+0.00067}_{-0.00095}$ & $-111.8$ & $87.84$ & 2.4343 \\ \hline
        2019-08-16 & $-120.94\,^{+20.81}_{-6.77}$ & $88.60\,^{+0.46}_{-0.51}$ & $2.4352\,^{+0.00011}_{-0.00007}$ & $-122.9$ & $88.60$ & 2.4352 \\ \hline
    \end{tabular}
    
    \label{tab:mcmc_sample_estimates}
\end{table*}

Figure \ref{fig:glint_map_results} displays the brightness maps created from the best fit MCMC parameters. The arrangement of mirrors into triplets is clearly seen across varying latitudes in the body fixed frame and the brightest regions of the surface reflectivity align well with the expected mirror positions. Table \ref{tab:mcmc_sample_estimates} displays the best estimates and errors calculated from taking the mode and FWHM of the sample distributions.

It should also be noted that for the purpose of creating a right-handed coordinate system, our spin-axis results would be inverted to represent the true orientation of Ajisai i.e. pointing towards the south celestial pole as opposed to the north.
\begin{figure}
    \centering
    \includegraphics
    [width = \linewidth]{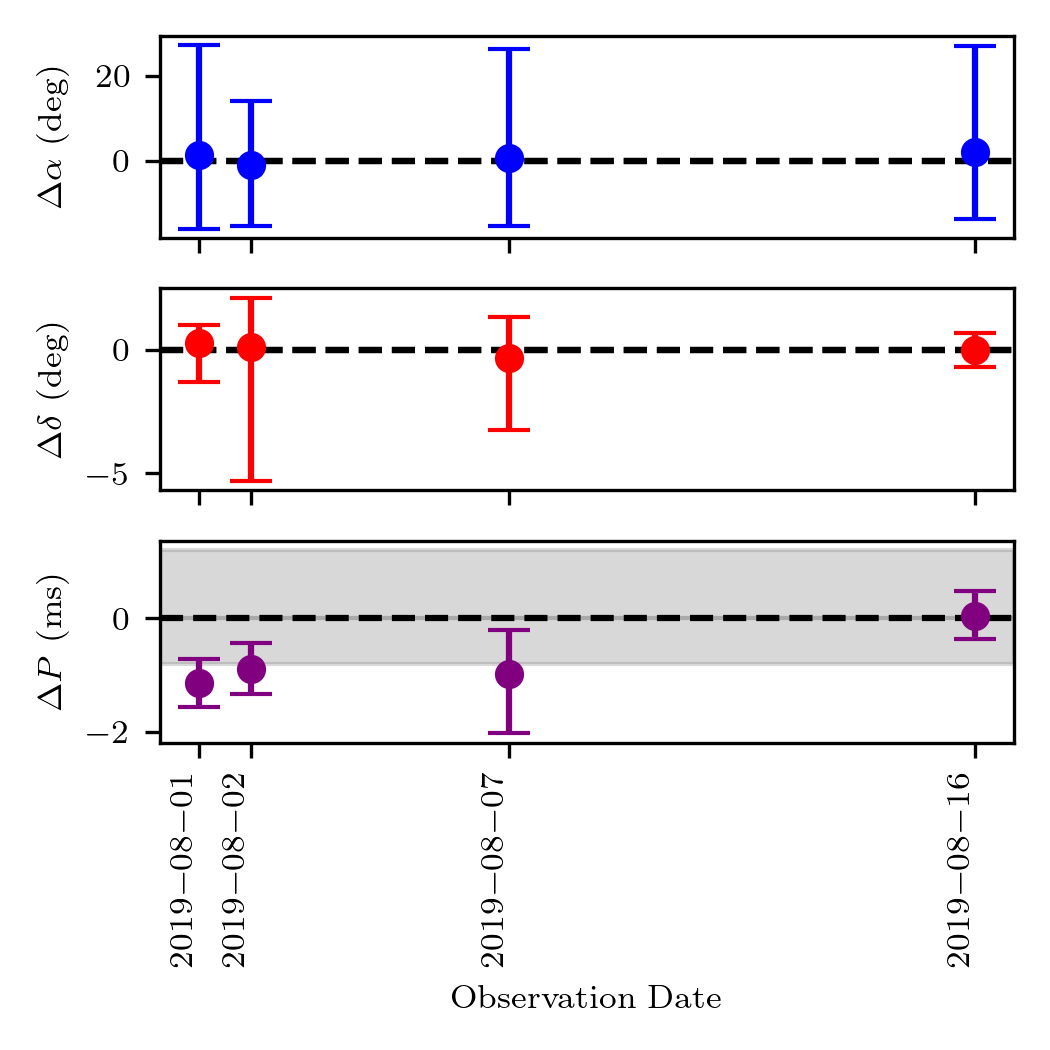}
    \caption{Residual measurements of parameters (estimate - empirical) for each observation of the Ajisai satellite. Error bars are calculated in quadrature from MCMC estimates and quoted residual/RMS measurements from empirical models. The grey shaded bounds illustrate the maximum and minimum variation around the empirical period seen due to changes in solar illumination duration in \protect\cite{2017AdSpR..60.1389K}}
    \label{fig:residual_measurements}
\end{figure}

Figure \ref{fig:residual_measurements} displays the residual measurements between the MCMC estimate and the empirically predicted values with the errors combined in quadrature. It is clear that the spin-pole estimations are close to the empirical value predictions and the first and fourth observation are particularly well constrained. The spin-pole $\delta$ residual for the second observation is not well constrained, however this is expected due to this observation covering fewer mirrors.

The empirical model provides a prediction of only the long-term, secular evolution of the spin period; it does not account for short-term variations, which could be driven---at least in part---by factors such as changes in solar illumination \citep{2017AdSpR..60.1389K} and asymmetry in the surface reflectivity introduced by the launch adapter ring and polar cap \citep{2019AdSpR..63...63H}. We find that only the 2019-08-16 observation agrees with the empirical model, but all four lie within the interval (-0.8 to 1.2 ms) reported by \cite{2017AdSpR..60.1389K}.

It is also worthwhile to mention that the relative motion between the observer and the target affects the measured apparent spin period regardless of the chosen reference frame—be it Earth-Centered Inertial (e.g., ICRF), Earth-Centered Earth-Fixed (e.g., ITRF), or the satellite’s body frame. We determine the rotation period by forward-modelling the ICRF phase-angle bisector into the body frame and maximizing its alignment with a static body-fixed mirror map; the fitted P is therefore the sidereal spin period, with observer/Sun relative motion handled inside the model rather than corrected afterwards. In contrast, \cite{CALATRONI2025}, first measure an apparent (synodic) period from inter-flash spacings within a triplet ("four-flash method"), and then apply a geometric correction to obtain the sidereal period.
\section{Summary}
We have demonstrated the feasibility of determining the spin-state of the Ajisai satellite through high-cadence photometric observations obtained using the SuperWASP instrument. We analysed four observations in August 2019, capturing sub-second glints consistent with specular reflections off Ajisai’s mirrors. By transforming TLE-derived vectors from the ICRF into the satellite’s body-fixed frame and generating latitude-longitude brightness maps, we showcase a novel method for mapping surface reflectivity.

Our methodology combines precision light curve extraction from streaked images with an MCMC-driven fitting process that aligns brightness maps in the satellite's body-fixed frame with modelled mirror positions. This indicates that constrained spin-state estimations can be derived from high-cadence photometry without relying on complex BRDF modelling or detailed prior knowledge of the satellite’s shape which is consistent with earlier photometric studies of Ajisai. The spin-state of a satellite significantly influences the morphology of light curves, governing the periodicity and modulation patterns that can be further exploited using machine learning models such as boosted decision trees \citep{2024RASTI...3..247S}.

Our best-fit spin-state solutions are consistent with empirical model predictions and observed variations in spin-period, affirming the adequacy of our mirror models in capturing the satellite's reflective behaviour. Notably, these results were achieved using relatively modest ground-based hardware adapted for high temporal resolution.

The original SuperWASP instrument used in this study has since been upgraded to STING, a wide-field multi-colour system that began routine operations in 2023 \citep{AIREY20255757}. This newer system, with reduced read noise (due to CMOS over CCD) and four distinct bandpasses, offers enhanced capabilities for resolving short-timescale glints with higher precision. In the case of Ajisai and similar objects, this will improve spin-state extraction and enable better resolution of closely separated objects during Rendezvous and Proximity Operations (RPO) or Active Debris Removal (ADR) missions.

These results affirm the value of high-cadence optical photometry for SDA applications, particularly for characterising the attitude of tumbling or rapidly spinning LEO objects—a critical capability in an increasingly congested orbital environment.

\section*{Acknowledgements}
This work makes use of data from the original SuperWASP instrument operated on the island of La Palma by the University of Warwick in the Spanish Observatory del Roque de los Muchachos of the Instituto de Astrofísica de Canarias. JAB acknowledges support from the Science and Technology Facilities Council
(grant ST/Y50998X/1). We thank Nikolay Koshkin at the Odessa National University for sharing their measurements of the Ajisai mirror positions. For the purpose of open access, the author has applied a Creative Commons Attribution (CC-BY) licence to any Author Accepted Manuscript version arising from this submission.

The authors would like to thank the reviewers for their comments and suggestions that helped to improve the quality of the final manuscript.

\section*{Conflict of Interest}
Authors declare no conflict of interest.

\section*{Data Availability}
The data underlying this article will be shared on reasonable request to the corresponding author.



\bibliographystyle{rasti}
\bibliography{example} 




\appendix
\section{Empirical Model Calculations}
\label{appendix:empirical_model_cals}
\subsection*{Spin-Axis Evolution}
The orientation of the Ajisai spin axis can be expressed by a set of time-dependent equations, following the empirical model developed by \cite{2016AdSpR..57..983K}. The coordinates of a vector oriented at the azimuthal angle \(\phi\) about the concentric cone of radius \(R\), and with its axis at right ascension \(\alpha\) and declination \(\delta\), are defined by the transformation:

\begin{equation}
S = R_3(-\alpha) \, R_2(\delta - 90^\circ) \, R_3(-\phi) \, R_2(-R),
\label{eq:spin_matrix}
\end{equation}

where \(R_3\) and \(R_2\) are the rotation matrices about the $z$- and $y$-axes, respectively, as defined in Equations~\ref{eq:R3-gamma} and~\ref{eq:R2}.

\noindent The inertial orientation of the nutation cone axis \(\vec{v}_N\) is calculated as:
\begin{equation}
\vec{v}_N = S \cdot \vec{z}, \quad \text{with } \vec{z} = [0, 0, 1]
\end{equation}

\noindent The parameters for the matrix \(S\) describing the precession cone are:
\begin{enumerate}

  \item Right ascension of the cone axis: \(\alpha = 88.90^\circ\)
  \item Declination of the cone axis: \(\delta = -88.85^\circ\)
  \item Radius of the precession cone: \(R = 1.08^\circ\)]
  \item Precession angle (azimuth): \(\phi = -0.0277404\,D + 452.803^\circ\)
\end{enumerate}
where \(D\) is the number of days since launch (MJD 46654.86).

\noindent The spin axis vector \(\vec{v}_S\) is calculated using a time-varying nutation cone defined about \(\vec{v}_N\), with parameters:
\begin{enumerate}
  \item Nutation cone radius:
  \[
  R = 7.66693 \times 10^{-9}\,D^2 - 0.0000656721\,D + 1.42878^\circ,
  \]
  \item Nutation angle (azimuth):
  \[
  \phi = 8.08453575 \times 10^{-7}\,D^2 + 3.07506\,D - 19238.5^\circ.
  \]
\end{enumerate}

\noindent Using these parameters, the transformation matrix \(S\) is recomputed with the updated \(\alpha\) and \(\delta\) (from \(\vec{v}_N\)), and the spin axis vector is given by:
\begin{equation}
\vec{v}_S = S \cdot \vec{z}.
\end{equation}

\subsection*{Spin Period Evolution}
The spin period \(P\), of Ajisai is modelled as an exponential trend function \citep{2017AdSpR..60.1389K} :
\begin{equation}
P(D) = 1.4934 \cdot \exp(0.000040553 \cdot D) \, \text{[s]},
\end{equation}
which describes the gradual loss of rotational energy due to weak magnetic torque effects.

\section{Separation Calculation}\label{appendix:sep_calc}
Our calculation of the quantity $sep(x_i,\,y_j)$ is calculated using the Haversine formula for small distances \citep{Sinnott:1984zz}. These are described by Equations \ref{eq:haver_1} and \ref{eq:haver_2} respectively. The subscripts of $\lambda$ and $\phi$ denote the longitude and latitude components of $x_i$ and $y_j$ respectively.

\begin{equation}
\begin{split}
A(x_i, y_j)
&= \sin^{2}\!\left(\frac{(y_j)_{\phi} - (x_i)_{\phi}}{2}\right) \\
&\quad + \cos\!\big((x_i)_{\phi}\big)\,\cos\!\big((y_j)_{\phi}\big)\,
    \sin^{2}\!\left(\frac{(y_j)_{\lambda} - (x_i)_{\lambda}}{2}\right)
\end{split}
\label{eq:haver_1}
\end{equation}

\begin{equation}
    sep(x_i,\,y_j) =
2 \, \arctan2\!\left(
      \sqrt{A(x_i,\,y_j)},
      \sqrt{1 - A(x_i,\,y_j)}
\right)
    \label{eq:haver_2}
\end{equation}

\bsp	
\label{lastpage}
\end{document}